\RequirePackage{fix-cm}
\documentclass[smallcondensed]{svjour3} 
\smartqed  
\usepackage{graphicx}
\usepackage{natbib}
\usepackage{xcolor}
\usepackage{amssymb}
\usepackage{amsmath}
\usepackage{lineno}
\usepackage{setspace}
\usepackage{hyperref}
\usepackage{longtable}
\usepackage{booktabs,subcaption,amsfonts,dcolumn}
\usepackage{algorithm}
\usepackage{algorithmic}
\usepackage{pifont}
\usepackage[shortlabels]{enumitem}
\usepackage{siunitx}
\usepackage{booktabs,colortbl,array}
\usepackage{pgfplotstable}
\pgfplotsset{compat=1.8}
\usepackage{multirow}
\usepackage{array}
\usepackage{hhline}
\usepackage{booktabs}
\usepackage{appendix}
\definecolor{rulecolor}{RGB}{0,71,171}
\definecolor{tableheadcolor}{gray}{0.92}

\journalname{Preprint submitted to ArXiv}

\begin{document}

\title{Modelling parametric uncertainty in PDEs models via Physics-Informed Neural Networks}

\titlerunning{Parametric uncertainty in PDEs via Physics-Informed Neural Networks}

\author{Milad Panahi \and
        Giovanni Michele Porta \and
         Monica Riva \and
        Alberto Guadagnini
}
\date{Received: August 08, 2024 / Accepted: date}
\institute{M. Panahi (0000-0002-8776-5297)\and G. M. Porta* (0000-0002-0636-373X) \and A. Guadagnini (0000-0003-3959-9690) \and M. Riva (0000-0002-7304-4114) \at Dipartimento di Ingegneria Civile e Ambientale, Politecnico di Milano, Piazza L. da Vinci 32, 20133 Milano, Italy \\
    * Corresponding author \email{giovanni.porta@polimi.it}
}

\maketitle
\begin{abstract}
We provide an approach enabling one to employ physics-informed neural networks (PINNs) for uncertainty quantification. Our approach is applicable to systems where observations are scarce (or even lacking), these being typical situations associated with subsurface water bodies. Our novel physics-informed neural network under uncertainty (PINN-UU) integrates the space-time domain across which processes take place and uncertain parameter spaces within a unique computational domain. PINN-UU is then trained to satisfy the relevant physical principles (e.g., mass conservation) in the defined input domain. We employ a stage training approach via transfer learning to accommodate high-dimensional solution spaces. We demonstrate the effectiveness of PINN-UU in a scenario associated with reactive transport in porous media, showcasing its reliability, efficiency, and applicability to sensitivity analysis. PINN-UU emerges as a promising tool for robust uncertainty quantification, with broad applicability to groundwater systems. As such, it can be considered as a valuable alternative to traditional methods such as multi-realization Monte Carlo simulations based on direct solvers or black-box surrogate models.

\section*{Article Highlights}
\begin{itemize}
    \item PINN-UU, a physics informed neural network is introduced to solve PDEs with uncertain physical parameters
    \item PINN-UU accurately reproduces direct solver results for multiple parameter realizations
    \item A stage learning approach is employed to overcome convergence failures in the optimization problem
\end{itemize}

\end{abstract}
\section{Introduction}
\label{section_introduction}

Groundwater is a critical component of the hydrologic cycle, hydraulic redistribution across the subsurface being a major driver supporting crops, vegetation and surface water bodies \citep{condon2021global}. Having at our disposal effective approaches to groundwater modeling can contribute to expand the boundaries of our knowledge of processes that can govern system dynamics at various scales and ultimately imprint the long-term behavior of hydrogeological scenarios of interest \citep{bethke2022geochemical}. Due to their inherent simplification of reality \citep{crawford1999geochemical}, models are typically subject to epistemic and aleatory uncertainty. Key challenges arise in conceptualizing the system behavior and providing an associated mathematical formulation, defining and characterizing input parameters, and addressing the associated uncertainties. The latter element is especially critical for high-dimensional parameter spaces (e.g., \cite{neuman2003maximum, tartakovsky2013assessment, ceriotti2018local, psaros2023uncertainty} and references therein).

Model parameter estimation is commonly carried out within an inverse problem framework \citep{carrera1986estimation}. Model calibration can be challenging and costly, particularly when dealing with subsurface systems where measurements are scarce and sparse (in space and time) or (sometimes) not available \citep{dai2004inverse}. To address these challenges, it is important to provide tools that can assist one to prioritize parameters to be estimated during characterization and optimize data acquisition during sampling campaigns \citep{ceriotti2018local, janetti2021natural}. Robust sensitivity analyses (SA) and forward uncertainty quantification (UQ) \citep{lee2009comparative} techniques are key to achieving these objectives.

In the realm of uncertainty quantification, Monte Carlo (MC) approaches have been broadly employed (e.g., \cite{ballio2004convergence,panzeri2014comparison} and references therein). These typically rely on multiple evaluations of the system model, each of these corresponding to a randomly sampled set of (uncertain) model input parameters. As such, the ensuing computational cost can be significant, depending on model complexity (in terms of processes included and parametrization). Reduced complexity approaches based on, e.g., analytical and semi-analytical \citep{pasetto2014reduced, pasetto2011pod, he2011enhanced} models, polynomial chaos expansions \citep{riva2015probabilistic, fajraoui2011use, janetti2021natural} and neural networks \citep{tang2020deep, tang2021deep, liu20213d, manzoni2023probabilistic}, are valuable to approximate the behavior of complex groundwater systems while significantly reducing computational costs \citep{asher2015review}. Each of these approaches offers unique advantages and limitations, and their applicability depends on the specific context under analysis.

Deep learning has emerged as a major tool in several fields of research. In this framework, deep neural networks (DNNs) are increasingly employed to tackle classical applied mathematics problems, such as to approximate the solution of partial differential equations (PDEs) \citep{Blechschmidt}. This shift in scientific computing is facilitated by the universal approximation capability and high expressivity of neural networks, making them particularly valuable for solving complex PDEs characterized by significant nonlinearities. DNNs are ansatz spaces for the solutions of PDEs \citep{mishra2022estimates} relying on collocating PDEs residual at training points \citep{cybenko1989approximation, barron1993universal, kingma20153rd}. Physics-Informed Neural Networks (PINNs) \citep{lagaris1998artificial, lagaris2000neural, lecun2015deep, raissi2019physics, raissi2017physics, raissi2020hidden} have recently emerged as an approximation technique, by transforming the task of directly discretizing differential operators into an optimization problem focused on minimizing a loss function. PINNs achieve this by integrating the mathematical model depicting the dynamics of the system considered into the neural network architecture and enhancing the loss function with a residual term derived from the governing equation (that are typically formulated upon resting on physical principles, e.g., mass, momentum or energy balances). This residual term serves as a constraining factor, effectively bounding the acceptable solution space.
In recent years, there has been an exponential growth in the amount of studies presenting algorithms with PINNs for various applications (e.g., \cite{raissi2018hidden, ray2018artificial, mishra2018machine, lye2021iterative}). PINNs have also been proposed in the context of groundwater modeling and have shown a remarkable potential in simulating the intricate dynamics of subsurface flow and transport.(e.g., \cite{he2021physics}, \cite{shen2023differentiable}, \cite{yeung2022physics}, \cite{tipireddy2022multistep}, and references therein).

In this work we propose a PINN-based method to address epistemic uncertainty associated with unknown physical parameters for sparse-to-none data situations. To address this challenge, we rely on an approach based on a physics-informed neural network under uncertainty (hereafter termed PINN-UU). The concept underlying the method rests on the definition of the computational domain where the problem solution is sought as comprising both ($a$) the spatiotemporal domain and ($b$) the space of uncertain parameters. In this sense, our method is inspired to intrusive UQ methods, where the solution accounts directly for stochastic variables (e.g., uncertain parameters \citep{LemaitreKnio}). While traditional intrusive UQ methods require major changes in problem formulation and algorithms (\cite{turnquist2019multiuq}), PINN-UU offers a framework to streamline the solution strategy. Our approach is not geared to parameter estimation, but rather to exploring the impact of parametric uncertainty onto the system outputs. Thus, we solely rely on the constraints of the physics embedded in the governing equations for solving PDEs (in the form of residual training), as opposed to approaches that combine physical laws and scattered (noisy) measurements for learning stochastic processes and addressing uncertainty quantification (e.g. MH-PINNs \citep{zou2023hydra}, and B-PINNs \citep{yang2021b}). To this end, we introduce a methodology to fine-tune the neural network-based solution ansatz via transfer learning. In the specific numerical examples discussed here our objective is to harness the capabilities of PINNs to provide an efficient and accurate tool for UQ in predicting solute transport in porous media. We thus showcase the application of PINN-UU to the solution of a scenario associated with solute transport in the presence of adsorption.

The study is organized as follows. In Section \ref{ProblemSet} we provide the overall formulation of our new abstract PDE setting and describe briefly vanilla PINNs as the base PDE solver for our new problem set. PINN-UU is described in Section \ref{PINN-UU}. Section \ref{results_section} focuses on the results of its application to an exemplary scenario associated with solute transport in the presence of sorption in a porous medium and on a comparison against an approach to the solution of the PDE governing the process analyzed through a typical finite difference method (FDM). We then analyze the applicability of PINN-UU within a global sensitivity analysis framework.
\\

\section{Methodology}\label{ProblemSet}
\subsection{Problem formulation}

\begin{table}
\begin{subtable}[t]{0.96\textwidth}
\centering
\caption{\footnotesize}
    \begin{tabular}{c||c}
      \toprule 
      \textbf{1) Forward stochastic} & Property\\
      \midrule
      $\mathcal{F}_{x,\lambda}$, $\mathcal{B}_{x,\lambda}$ & known\\\\
      $f_{\boldsymbol{\lambda}}(\mathbf{x}), b_{\boldsymbol{\lambda}}(\mathbf{x})$ & known\\\\
      $u(\mathbf{x};\boldsymbol{\lambda})$ & unknown\\\\
      $\boldsymbol{\lambda}$ & stochastic\\\\
      Objective & find $u(\mathbf{x};\boldsymbol{\lambda})$ for $\forall \mathbf{x} \in \Omega_{\boldsymbol{\lambda}}$\\
      \bottomrule 
    \end{tabular}
\label{tab:table1_a}
\end{subtable}

\bigskip 
\begin{subtable}[t]{0.96\textwidth}
\centering
\caption{\footnotesize}
    \begin{tabular}{c||c}
      \toprule 
      \textbf{2) Forward deterministic} & Property\\
      \midrule
      $\mathcal{F}_{x}$, $\mathcal{B}_{x}$ & known\\\\
      $f(\mathbf{x}), b(\mathbf{x})$ & known \\\\
      $u(\mathbf{x})$ & unknown\\\\
      $\boldsymbol{\lambda}$ & deterministic\\\\
       Objective & find $u(\mathbf{x})$ for $\forall \mathbf{x} \in \Omega$\\
      \bottomrule 
    \end{tabular}
\label{tab:table1_b}
\end{subtable}
\hspace{\fill}
\caption{\footnotesize The (a) forward stochastic scenario properties addressed in this work compared against the classical (b) forward deterministic scenario.
}
\label{tab:table1}
\end{table}

Consider the following partial differential equation (PDE) describing the space-time dynamics of the state variable, $u(\mathbf{x};\boldsymbol{\lambda})$, of a physical system

\begin{equation}
    \mathcal{F}_{x,\lambda}[u(\mathbf{x};\boldsymbol{\lambda})] = f(\mathbf{x};\boldsymbol{\lambda}) \quad \mathbf{x} \in \Omega,
\label{PDE}
\end{equation}
where $\mathbf{x}$ denotes the $D_{x}$-dimensional \textbf{space-time} coordinate, $\boldsymbol{\lambda} \in \Lambda \subseteq \mathbb{R}^{N_\lambda}$ represents a vector of (generally unknown) physical/model parameters, $F_{x,\lambda}$ (hereafter referred to as $F_{\boldsymbol{\lambda}}$ for brevity of notation) denotes the differential operator and $f(\mathbf{x};\boldsymbol{\lambda})$ (hereafter referred to as $f_{\boldsymbol{\lambda}}$) is the forcing function. Equation\eqref{PDE} is subject to the following boundary/initial conditions \addtocounter{equation}{-1}
\begin{subequations}
    \begin{align}
        \mathcal{B}_{x,\lambda}[u(\mathbf{x};\boldsymbol{\lambda})] & =b(\mathbf{x};\boldsymbol{\lambda}) \quad \mathbf{x} \in \Gamma,\label{eq:diffusionpartialdiff_boundaries0}
    \end{align}
\end{subequations}
where $B_{x,\lambda}$ (hereafter referred to as $B_{\boldsymbol{\lambda}}$) denotes the boundary/initial operators and $b(\mathbf{x};\boldsymbol{\lambda})$ (hereafter referred to as $b_{\boldsymbol{\lambda}}$) is the boundary/initial function.

Physical parameters are included in vector $\boldsymbol{\lambda} = [\boldsymbol{\lambda}_k, \boldsymbol{\lambda}_u]$. The latter is formed by $N_{\boldsymbol{\lambda}_{k}}$ deterministic ($\boldsymbol{\lambda}_{k}$) and $N_{\boldsymbol{\lambda}_{u}}$ uncertain ($\boldsymbol{\lambda}_{u}$) parameters. We consider all entries of $\boldsymbol{\lambda}_{u}$ to be multivariate uniform stochastic variables, i.e., $\boldsymbol{\lambda}_{u} \sim \mathcal{U}(\boldsymbol{\lambda}_{u,min},\boldsymbol{\lambda}_{u,max})$, where $\boldsymbol{\lambda}_{u,min}$, $\boldsymbol{\lambda}_{u,max}$ represent lower and upper bounds of the parameter supports, respectively. The parameter space $\Lambda$ identifies the domain considering all possible parameters combinations. 
We introduce the problem domain $\Omega_{\boldsymbol{\lambda}} = \Lambda \times \Omega$, and $\Gamma_{\boldsymbol{\lambda}} = \Lambda \times \Gamma$ by jointly considering the spatiotemporal domain ($\Omega$) and domain boundary $\Gamma$ together with the parameter support space ($\Lambda$). Thus, the dimension of $\Omega_{\boldsymbol{\lambda}}$ corresponds to $N_{\Bar{P}}= D_x + N_{\boldsymbol{\lambda}_{u}}$. Our objective is to obtain the solution $u(\mathbf{x};\boldsymbol{\lambda})$ at every $(\mathbf{x};\boldsymbol{\lambda}) \in \Omega_{\boldsymbol{\lambda}}$.

Following \cite{psaros2023uncertainty}, we classify problem sets as neural PDEs when we have complete knowledge of the operators $F_{\boldsymbol{\lambda}}$ and $B_{\boldsymbol{\lambda}}$. In this context, the distinctive element of neural PDEs used in our work is their dependence on neural networks (NNs) as a central computational framework. Our methodological workflow is then characterized by the key features listed in Table \ref{tab:table1_a} and pertains to a \textit{forward stochastic} PDE problem (i.e., $F_{\boldsymbol{\lambda}}$ and $B_{\boldsymbol{\lambda}}$ are stochastic) as opposed to conventional \textit{forward deterministic} PDEs, where $F_{\boldsymbol{\lambda}}$ and $B_{\boldsymbol{\lambda}}$ are deterministic (see Table \ref{tab:table1_b}).

\begin{figure*}
	\centering
	\includegraphics[width=1\textwidth]{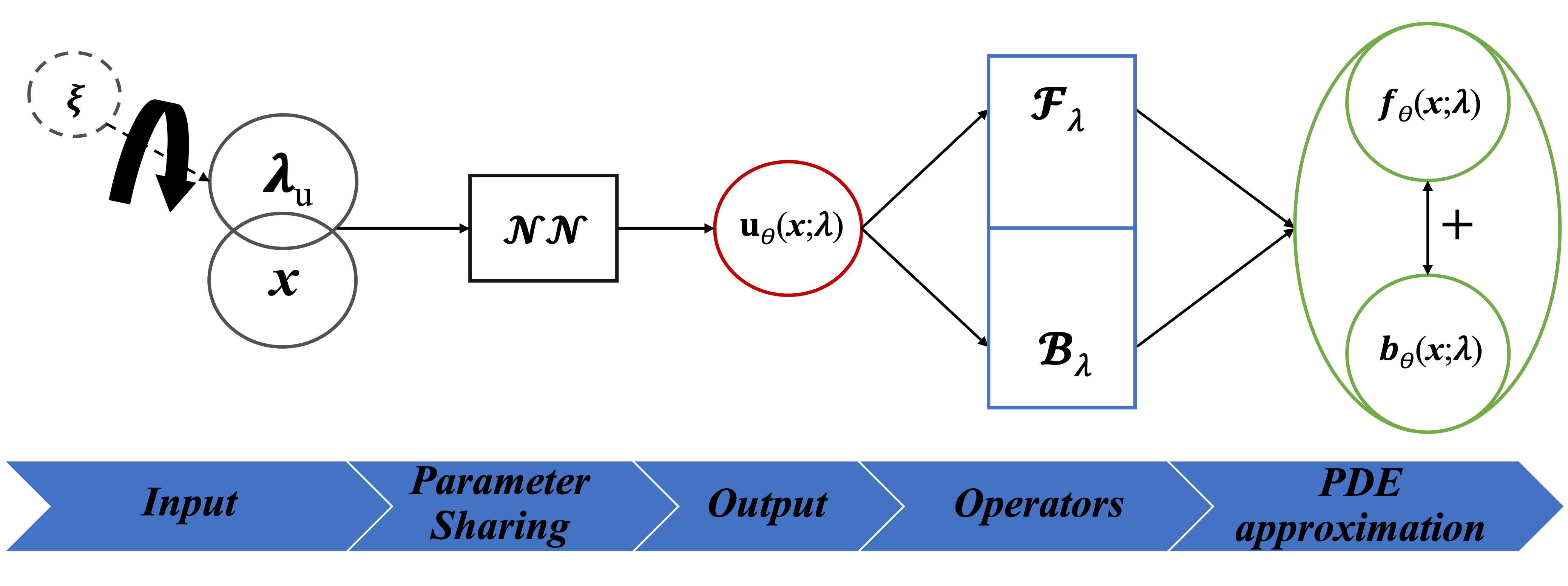}
	\caption{PINN-UU architecture. Here, $\xi$ denotes a random number generator associated with a desired distribution (e.g., uniform) and employed to sample training points in the $\Omega_{\boldsymbol{\lambda}}$ space.}
	\label{PINN-UU-scheme}
\end{figure*}

\begin{algorithm}
    \caption{Training PINN-UU}\label{alg2}
    \begin{algorithmic}
    \STATE \textbf{Inputs}: 
    \begin{enumerate}[i.]
        \item Multi-dimensional domain $\Omega_{\boldsymbol{\lambda}}$ 
        
        \item differential operator $\mathcal{F_{\boldsymbol{\lambda}}}$, boundary/initial operator $\mathcal{B_{\boldsymbol{\lambda}}}$ and their corresponding source terms $f_{\boldsymbol{\lambda}}$ and $b_{\boldsymbol{\lambda}}$ for PDE \eqref{PDE}
    \end{enumerate}

    \bigskip

    \STATE \textbf{Goal}: Find PINN-UU solution $u^{*} = u_{\theta_{\lambda}^{*}}$ for approximating PDE \eqref{PDE}.
    \bigskip
    \newcommand{\norm}[1]{\lvert #1 \rvert}
    \STATE \textbf{Step 1}: Construct the neural network $u_\theta$ architecture, specifying the input dimension as $N_{\Bar{P}}= D_x + N_{\boldsymbol{\lambda}_{u}}$

    \bigskip
    \STATE \textbf{Step 2}: $k \gets N_{\boldsymbol{\lambda}_{u}}$, assign $N_{\boldsymbol{\lambda}_{u}}$ to the iterator index $k$, and initialize the neural network parameter vector $\Bar{\theta}^k_{\boldsymbol{\lambda}}$ using Xavier Glorot's initialization.
    \bigskip
    \WHILE{$k \geq 0$}
    
    \bigskip
    \STATE \textbf{Step 3}: Set the first $k$ entries of $\{\lambda^i_{u}\}_{i=1}^{k}$ equal to the \textbf{mean} of their corresponding \textbf{probability distribution}, and sample the remaining $N_{\boldsymbol{\lambda}_{u}} - k$ members, from their corresponding \textbf{probability distribution}.
    
    \bigskip
    \STATE \textbf{Step 4}: For the initialized value of the network parameter vector $\Bar{\theta}^k_{\boldsymbol{\lambda}}$, run an optimization algorithm of choice until an approximate local minimum $\theta_{\boldsymbol{\lambda}}^{k*}$ of Eq. \eqref{PINNLoss1} is reached.
    \bigskip
    \STATE \textbf{Step 5}: $\Bar{\theta}^{k-1}_{\boldsymbol{\lambda}} \leftarrow \theta_{\boldsymbol{\lambda}}^{k*}$, i.e., we transfer the learned parameters of previous stage $\theta_{\boldsymbol{\lambda}}^{k*}$ and use it to initialize the neural parameters $\Bar{\theta}^{k-1}_{\boldsymbol{\lambda}}$ for the next step ($k-1$)
    \bigskip
    \STATE \textbf{Step 6}: $k \gets k - 1$  iterate over while loop, i.e., decrease the iterator index by one
    \bigskip
    \ENDWHILE
    \bigskip
    \STATE \textbf{Step 7}: The map $u^{*} = u_{\theta_{\boldsymbol{\lambda}}^{k*}}|_{k=0}$ is the desired PINN-UU for the approximation of the solution $u(\mathbf{x};\boldsymbol{\lambda})$ of PDE \eqref{PDE}
    \bigskip
    \end{algorithmic}
\end{algorithm}

\section{Physics Informed Neural Network Under Uncertainty (PINN-UU)}\label{PINN-UU}
In the following we describe the \textbf{PINN} \textbf{U}nder \textbf{U}ncertainty (PINN-UU). Our approach is built on classical PINNs. A schematic depiction of our PINN-UU is shown in Fig. \ref{PINN-UU-scheme}. The dependency of the network parameters on the uncertain (stochastic) parameters $\boldsymbol{\lambda} u \in \boldsymbol{\lambda}$ is fully taken into account, as our network receives the PDE parameter values in the input stage.

The PINN-UU network comprises an input layer that incorporates the parameter values, followed by multiple shared hidden layers with dropout regularization, and an output layer for the PDE unknown solution. Similar to PINNs, our neural network is also trained using a combined loss function that includes the residual of the PDE equations and the boundary/initial conditions.
In the context of our framework, the optimization problem to find neural networks $u_{\theta_{\boldsymbol{\lambda}}}$ with tuning parameters $\theta_{\boldsymbol{\lambda}}$ can be rewritten as:

\begin{equation}
\begin{aligned}
\hat{\theta}_{\boldsymbol{\lambda}}=\underset{\theta_{\boldsymbol{\lambda}}}{\operatorname{argmin}} \mathcal{L}(\theta_{\boldsymbol{\lambda}}), \quad \mathcal{L}(\theta_{\boldsymbol{\lambda}}) := &\int_{\Omega_{\boldsymbol{\lambda}}}\left|\mathcal{F}_{\boldsymbol{\lambda}}(u_{\theta_{\boldsymbol{\lambda}}}) - f_{\boldsymbol{\lambda}}\right|^2 \mathrm{~d} x + \\ &\int_{\Gamma_{\boldsymbol{\lambda}}}\left|\mathcal{B}_{\boldsymbol{\lambda}}(u_{\theta_{\boldsymbol{\lambda}}}) - b_{\boldsymbol{\lambda}}\right|^2 \mathrm{~d} x,
\end{aligned}
\end{equation}\label{PINNLoss1}
knowing that PDE \eqref{PDE} must hold for any $\mathbf{x} \in \Omega_{\boldsymbol{\lambda}}, \Gamma_{\boldsymbol{\lambda}}$. The training set (of size $N_f$) for the interior domain will include ($a$) $N_{\Bar{P}}$-tuplets of residual training points in $\Omega_{\boldsymbol{\lambda}}$ and ($b$) the training set of size $N_b$ and $N_0$ for the spatial boundary and initial conditions, respectively. The training strategy is presented in Algorithm \ref{alg2} and described in details in Section \ref{Transfer learning}. The neural network architecture is fixed throughout the training.

The minimization problem in Eq. \eqref{PINNLoss1} amounts to finding a minimum of a possibly non-convex function over a high-dimensional space. We solve this minimization problem with a first-order stochastic gradient descent, ADAM \citep{kingma2014adam}, followed by a higher-order optimization method, LBFGS \citep{liu1989limited}. Following determination of $\hat{\theta}_{\boldsymbol{\lambda}}$ as our best approximate/local minimum, the underlying network $u_\theta$ can be evaluated at any $\mathbf{x} \in \Omega_{\boldsymbol{\lambda}}$ as our PINN approximation for the solution $u$ of PDE \eqref{PDE}.

\subsection{Transfer learning (Warm-starting)}\label{Transfer learning}
The neural network parameter space grows with the dimension of problem sampling space ($\Omega_{\boldsymbol{\lambda}}$). This leads to a corresponding increase of computational complexity and memory requirements. This issue hampers the efficient training of PINNs and often results in slow convergence rate and increased likelihood of trapping in local minima. Introducing the PDE parameters as inputs to the PINNs has two main drawbacks: a) dealing with high-dimensional space leads to computational issues in the training phase; and b) the level of complexity of the loss surface increases, thus causing network optimization to fail or to require unaffordable computational resources. In order to alleviate these negative aspects, we rely on a broadly known technique termed transfer learning. Here, we train our PINN-UU via a stage learning strategy, where we initiate the training process by employing parameters retained from prior sessions (see Algorithm \ref{alg2}, steps 3 to 5), stored in memory, thereby adopting a \textit{warm-start} strategy. This is in contrast with the \textit{cold-start} approach, where training begins with entirely random parameters \citep{raissi2023open}. Doing so can speed up convergence of the optimization problem and minimizes the probability of being trapped in local minima. We start from lower dimensions (i.e., corresponding to the classic spatiotemporal domain with fixed parameter values as inputs) and use the trained network parameters as a first guess candidate to initialize training on higher dimensions as we introduce stochastic parameters one by one to the sampling space. In the first iteration all parameters in $\boldsymbol{\lambda}$ are considered deterministic and fixed to the mean value of their corresponding probability density function. We then introduce one stochastic parameter at each step according to the while loop shown in Algorithm \ref{alg2}). In other words, parameters are initially fixed to their mean value and then progressively sampled from their corresponding distributions.
According to \cite{mishra2023estimates}, our network generalization error can be upper-bounded and determined by training error and number of residual training points. Therefore, this network can (in principle) be a viable candidate as a neural solver to address our new problem formulation.

\section{Results and Discussion}\label{results_section}
\subsection{Problem formulation}\label{solute transport}
In this section we illustrate the results of numerical experiments aimed at examining the performance of the proposed PINN-UU. We investigate solute transport coupled with sorption in porous media. This is modeled upon relying on the Advection-Dispersion Equation (ADE) and including a retardation factor that is represented through the Langmuir sorption isotherm \citep{fetter1999contaminant}.

After describing the PDE formulation and the parameter settings in Section \ref{Governing Equation}, Section \ref{NN} summarizes the outcomes of our systematic exploration aimed at identifying the optimal network configuration. We then present in Section \ref{Error analysis} a comparison between the performance of PINN-UU and a reference Finite Difference Method (FDM). We consider an instance-wise (Section \ref{Instance wise}) level of comparison and a comparison based on the probability density function of model outputs (Section \ref{Statistical wise}). For the latter analysis, we rely on a large dataset comprising $10^6$ Monte Carlo realizations of the uncertain parameters $\boldsymbol{\lambda}_u$ and their corresponding PDE solution for PINN-UU and FDM, denoted in the following as $u_{pre}$ and $u_{ref}$, respectively. Finally, in Section \ref{SA} we illustrate the way PINN-UU can be employed in the context of global/local sensitivity analysis.\\\\

\subsubsection{Governing Equation}\label{Governing Equation}
The one-dimensional advection-dispersion equation coupled with Langmuir sorption isotherm can be expressed as:

\begin{flalign}\label{ADS PDE}
\left\{\begin{aligned}
& u_t\left[1+\frac{\rho_b}{\theta}\left(\frac{K_l Q}{\left(1+K_l u^2\right)}\right)\right]-\left(D u_{xx} - v u_x\right) = 0 \quad\quad x \in \Omega ;  \boldsymbol{\lambda}_u = [D, K_l, Q]\\
& v u - D u_x = Q_{in} \quad x \in \partial \Omega_F\\
& u_x = g_N \quad x \in \partial \Omega_N\\
& u = u_0 \quad x \in \partial\Omega_0,
\end{aligned}\right.
\end{flalign}
\begin{figure*}
	\centering
	\includegraphics[width=1\textwidth]{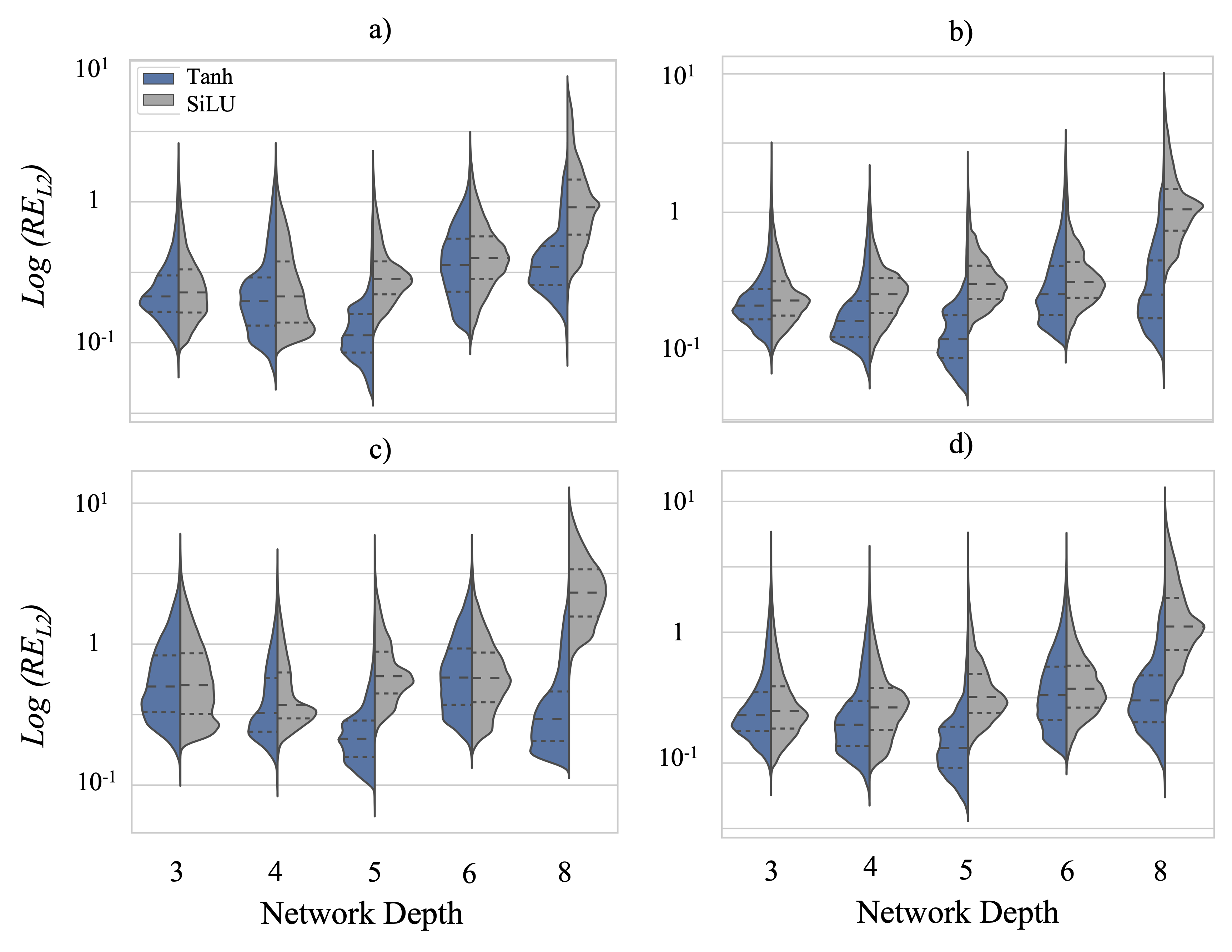}
	\caption{Model tuning: distribution of the relative percentage errors $RE_{L2}$ (Eq. \eqref{RE}) between the reference FDM solution and the examined PINN-UU architectures. Shaded areas with different colors, blue and grey, correspond to the two activation functions employed, i.e., Tanh and SiLU, respectively. Results are shown for three different regimes corresponding to a) low, b) medium, and c) high P\'eclet, as well as d) overall. Dashed lines correspond to the first, second, and third quartiles. The loosely dashed line and the densely dashed lines represent the median, and $(25^{th},75^{th})$ percentiles, respectively.}
	\label{Model_Selection}
\end{figure*}
where $u$ denotes concentration of the solute in the liquid phase in the spatiotemporal domain $\Omega$; $\mathbf{x} = (x,t)$ is space-time coordinate vector; $\rho_b$ and $\theta$ are bulk density and porosity of the porous medium, respectively; $K_l$ is the Langmuir adsorption constant (or adsorption rate); $Q$ is the maximum amount of solute that can be adsorbed per unit of solid surface (hereafter termed as maximum adsorption); $D$ is dispersion coefficient, embedding molecular diffusion and hydrodynamic dispersion; $v$ is the average pore velocity; $g_N$ is prescribed mass flux at the effluent boundary $\partial \Omega_N$; $Q_{in}$ is the solute mass flux imposed at the inlet boundary $\partial \Omega_F$; and $u_0$ is the initial condition for concentration, $\Omega_0$ denoting initial time.

Table \ref{table2} lists the parameter values/ranges that we employ to solve Eq. \eqref{ADS PDE} for the entire 5-dimensional space $\Omega_\lambda$. Our problem set mimics a continuous injection of a solute in a homogeneous porous domain where the initial concentration is set to zero and closely resembles conditions in column flow-through experiments. Each uncertain parameter is characterized by a uniform probability density function. The widths of the corresponding supports correspond to about an order of magnitude (in terms of parameter values) and are designed to include ranges of parameter values commonly employed and discussed in the literature. The P\'eclet number, $Pe$, resulting from our selected interval spans the range $(10,100)$, thus considering transition from mild to strongly advection-dominated conditions.

Average values of parameters $K_l$ and $Q$ are selected to yield a retardation factor approximately equal to $2$ when considering the reference concentration $u = Q_{in}/v = 1$.\\
\begin{table}[t]
\centering
\renewcommand{\arraystretch}{1.2}
\begin{tabular}{c|c|c}
\toprule
\textbf{Parameter} & \textbf{Ranges/Values} & \textbf{Variable Type}\\
\midrule
$(x[L] , t[T])$ or $\mathbf{x}$ & $[0,X = 30] \times [0,T = 15]$ &\\
$\lambda_1$ or $D[L^2T^{-1}]$ & $(3,30)$ &$\boldsymbol{\lambda}_u$\\
$\lambda_2$ or $K_l[L^3M^{-1}]$ & $(0.01, 0.1)$ &\\
$\lambda_3$ or $Q[MM^{-1}]$ & $(1,10)$ &\\
\midrule
\multirow{4}{*}{$Pe[-] = \frac{X \times v}{D} = \frac{30 \times 10}{D}$} & \multirow{1}{*}{$(10 , 100)$} & \multirow{4}{*}{--}\\
& $Pe < 15 \Rightarrow \text{Low}$ &\\
& $15 \leq Pe \leq 40 \Rightarrow \text{Med}$ &\\
& $Pe > 40 \Rightarrow \text{High}$ &\\
\midrule
$\rho_b[MM^{-3}]$ & $1.6$ &\\
$\theta[-]$ & $0.37$ &\\
$v[LT^{-1}]$ & $10$ &$\boldsymbol{\lambda}_k$\\
$Q_{in}[ML^{-2}T^{-1}]$ & $10$ &\\
$g_N[ML^{-1}]$ & $0$ &\\
$u_0[ML^{-3}]$ & $0$ &\\
\bottomrule
\end{tabular}
\caption{Description of problem setting. Training points are samples from $(\mathbf{x}, \boldsymbol{\lambda}_u)$ across our $5-dimensional$ space $\Omega_{\boldsymbol{\lambda}_u}$. Parameters $[D,K_l,Q]$ are categorized as uncertain, while the remaining parameters are fixed (i.e., are of $\boldsymbol{\lambda}_k$-type).}
\label{table2}
\end{table}

\subsubsection{Neural Network Architecture and Training Procedure}\label{NN}
We employ neural networks with diverse architectures to assess their ability to provide accurate solutions of Eq. \eqref{PDE}. 
The input layer comprises $5$ neurons (corresponding to our $5-dimensional$ space $\Omega_{\boldsymbol{\lambda}}$) to sample the residual training points. We then systematically vary the network depth (i.e., the number of hidden layers) and width (i.e., the number of neurons in each layer), to appraise different levels of complexity within the PDE solution space. Each of these configurations is associated with two activation functions, namely hyperbolic Tangent (Tanh) and Scaled Exponential Linear Unit (SiLU; also known as Swish). These activation functions are selected to introduce non-linearity into the network. The explored architectures include configurations with [depth, width] of $[3, 250]$, $[4, 220]$, $[5, 150]$, $[6, 50]$, and $[8, 100]$. Therefore, we explore a total of $10$ distinct network architectures.
\begin{table}[t]
\centering
\renewcommand{\arraystretch}{1.2}
\begin{tabular}{c|c|c}
\toprule
\textbf{Step ID ($N_{\boldsymbol{\lambda}_{u}} – k$)} & \textbf{$\Omega_{\boldsymbol{\lambda}}$} & \textbf{$[N_f,N_b,N_0]$}\\
\midrule
$0$ & $x \times t \times \bar{D} \times \bar{K_l} \times \bar{Q}$ & [$10k$, $50$, $50$]\\
\midrule
$1$ & $x \times t \times D \times \bar{K_l} \times \bar{Q}$ & [$200k$, $5k$, $5k$]\\
\midrule
$2$  & $x \times t \times D \times K_l \times \bar{Q}$  & [$1000k$, $50k$, $50k$]\\
\midrule
$3$  & $x \times t \times D \times K_l \times Q$ & [$3000k$, $100k$, $100k$]\\
\bottomrule
\end{tabular}
\caption{Training points sampling scheme; $k$ corresponds to $\times 10^3$; overbar symbols $\bar{\cdot}$ indicate parameters that are sampled equal to their corresponding distribution average value in each training iteration, as cited in Algorithm \ref{alg2}.}
\label{table3}
\end{table}

To enhance the network resilience and generalization ability, we add dropout layers with a dropout rate of $0.5$ after each hidden layer.
Adoption of, $(a)$ Glorot normal initialization \citep{glorot2010understanding} for the network weight parameters, and $(b)$ Soft Attention Mechanism (SAM) refined through adversarial learning strategy \citep{mcclenny2020self}, contributes to achieve stable training and to force the neural network to fit better stubborn spots (where the solution space includes regions with rapid variation due to stiffness) in the solution of our PDE, respectively. We also use mini-batch learning for PDE residual training points ($N_f$) to alleviate computational costs of large matrix multiplications, while keeping full-batch learning for initial and boundary residual training points ($N_0$ and $N_b$, respectively).

The training procedure is implemented in Tensorflow $2.10$ for $100k$, $50k$, and $10k$ iterations of ADAM for $N_{\boldsymbol{\lambda}_{u}} = 3$ steps of our multi stage training, consistent with the number of uncertain physical parameters ($D,K_l,Q$). This is then followed by $1k$ iterations of L-BFGS quasi-newton method in some cases to fine tune the network weights. The number of residual training points selected for the trials is also listed in Table \ref{table3}. We rely on the Sobol sampling strategy to optimize placement of training points across the domain.

Training is implemented on a single Nvidia GeForce RTX $4090$ GPU. We implement SAM and apply locally (i.e., per training point) adaptive loss weights initialized to be $1$. The exponentially decaying learning rates for network trainable parameters and self-adaptive weights are initialized at 4e-3 and 2e-3, respectively, with decay steps of $5\%$ of the total training iterations.

\begin{table}[t]
\centering
\renewcommand{\arraystretch}{1.2}
\begin{tabular}{c|c|c}
\toprule
\textbf{Transport Regime} & \textbf{Best Performance} & \textbf{Worst Performance}\\
\midrule
Low $Pe$ & [$0.038,13,0.067,3.78$] & [$1.23,10,0.1,1$]\\
\midrule
Med $Pe$  & [$0.043,18,0.03,1.64$] & [$1.64,37,0.01,10$]\\
\midrule
High $Pe$  & [$0.065,43,0.034,1.4$] & [$2.12,79,0.01,10$]\\
\bottomrule
\end{tabular}
\caption{Combinations of the transport model unknown parameters with their corresponding $RE_{L2}$ (Eq. \eqref{RE}) value for best and worst performance associated with the three distinct transport regimes, corresponding to low, medium, and high P\'eclet value. Values inside brackets correspond to $[RE_{L2},Pe,K_l,Q]$.}
\label{table4}
\end{table}

\begin{figure*}
	\centering
	\includegraphics[width=1\textwidth]{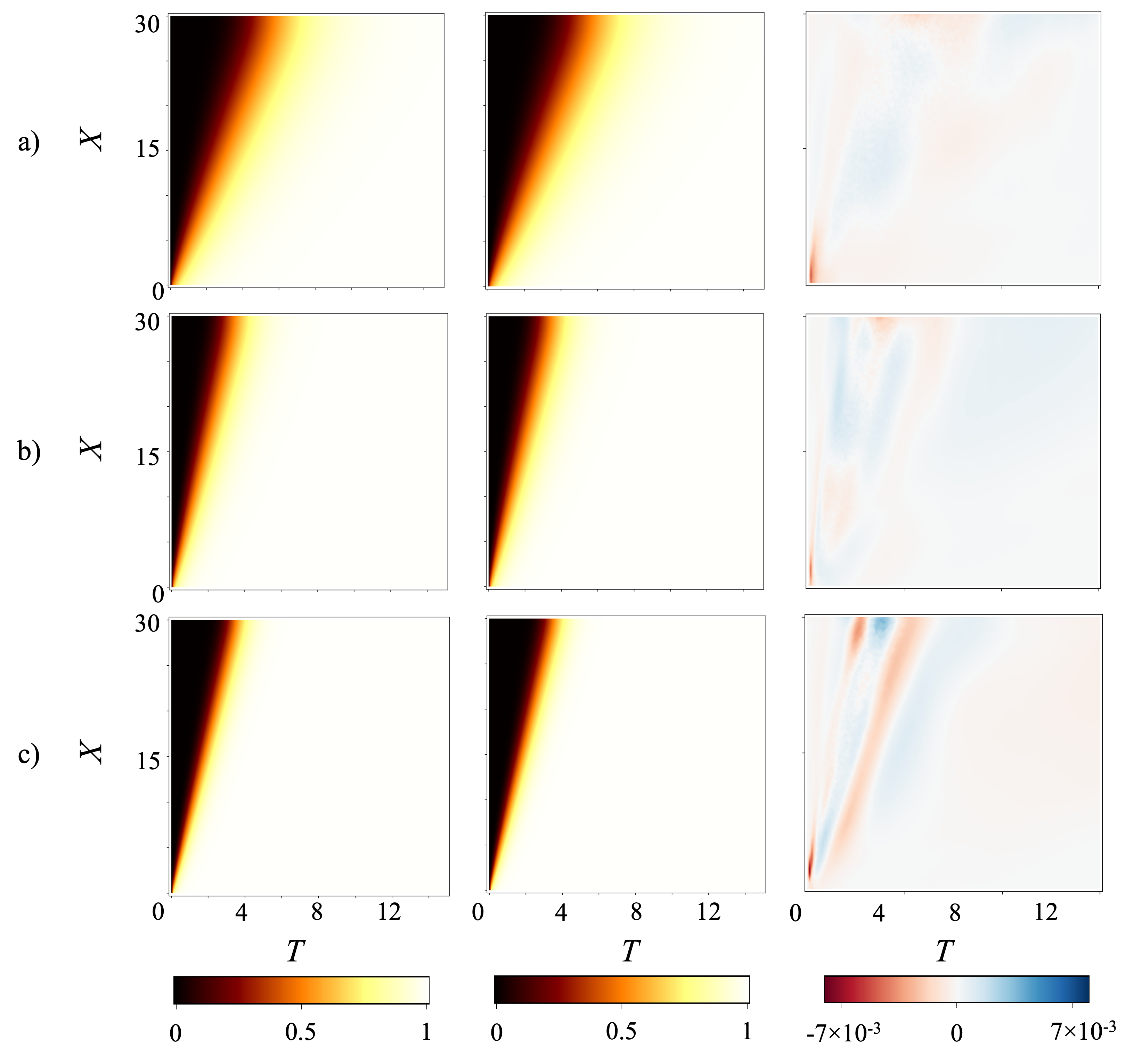}
	\caption{Space-time distribution of the results of the best performing PINN-UU, $u_{pre}$ ($1^\text{st}$ column), reference Finite Difference solution, $u_{ref}$ ($2^\text{nd}$ column), and their corresponding point-wise error [$u_{pre} - u_{ref}$] ($3^\text{rd}$ column), for a) low, b) medium, and c) high P\'eclet.}
	\label{Best Instance Wise}
\end{figure*}
Figure \ref{Model_Selection} depicts (sample) probability density functions (PDFs) of L2-Norm relative error percentages $RE_{L2}$, evaluated as:
\begin{equation}
RE_{L2} = 100\times\frac{\|u_{pre} - u_{ref}\|_2}{\|u_{ref}\|_2} \quad [\%],
\label{RE}
\end{equation}

The PDFs shown in Figure \ref{Model_Selection} are evaluated considering a Monte Carlo sample of $10^6$ realizations. it can be noted that similar PDFs of $RE_{L2}$ are obtained in all of the tested network configurations. Nevertheless, our results indicate that, in general, a network architecture featuring a depth of $5$ hidden layers and a width of $150$ neurons, trained with the Tanh activation function exhibits a slightly better performance (due to lower mean relative error and more precision) than the other configurations. In scenarios characterized by high P\'eclet numbers, leading to steep solution surfaces, the network $[5, 150]$ gives a performance similar to the high complexity network architecture with the deepest configuration, corresponding to network $[8, 100]$. We also observe a noteworthy distinction in the performance of the SiLU activation function when compared to Tanh, particularly in the context of deeper networks. On average, the SiLU activation function exhibits a slightly inferior performance, a phenomenon that may be attributed to challenges associated with vanishing gradients \citep{nwankpa2018activation}.\\\\
Based on our findings, selecting a moderately-deep PINN-UU, i.e., $[5, 150]$ as the most performant model yields accurate results within an acceptable computational time-frame of approximately $5$ GPU-Hours for the test case considered.

\begin{figure*}
	\centering
	\includegraphics[width=1\textwidth]{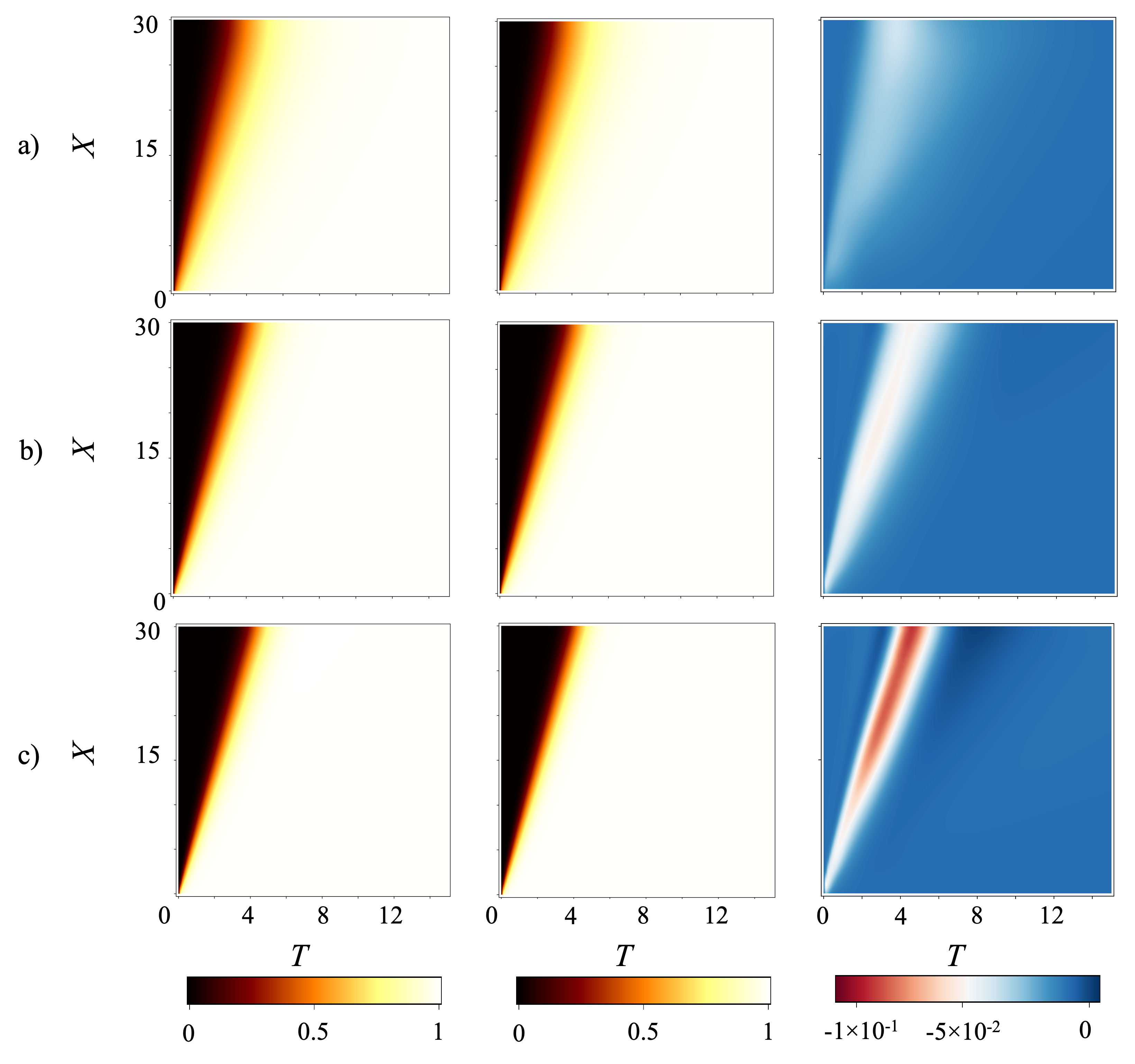}
	\caption{ Space-time distribution of the results of the worst performing PINN-UU ($1^\text{st}$ column), to reference Finite Difference solution ($2^\text{nd}$ column), and their corresponding point-wise error ($3^\text{rd}$ column) for a) low, b) medium, and c) high P\'eclet.}
	\label{Worst Instance Wise}
\end{figure*}

\subsection{Error Analysis}\label{Error analysis}
In this section, we perform a comprehensive analysis of the performance of the selected neural network model configuration (i.e., $[5, 150]$). We consider the results of individual realizations (Section \ref{Instance wise}) and an analysis of the ensuing probability density functions (Section \ref{Statistical wise}).

\subsubsection{Instance Wise Analysis}\label{Instance wise}
We perform here pairwise comparisons between the results obtained through PINN-UU and the reference solution (FDM) for selected parameter realizations. Figure \ref{Best Instance Wise} and Figure \ref{Worst Instance Wise} respectively illustrate the best- and worst-performing model realizations generated by PINN-UU (here denoted as $u_{pre}$) alongside the reference solutions (here denoted as $u_{ref}$) for three distinct transport regimes (low, medium, and high P\'eclet) as well as the space-time distribution of the related point-wise errors (calculated as $(u_{pre} - u_{ref})$). The associated $RE_{L2}$ value and parameter values related to best and worst perfomance are included in Table \ref{table4}.

Our results show that the largest point-wise errors are of the order of $0.1\%$ (for the best-performing realizations) and $10\%$ (for the worst-performing realizations). The regions around steep gradients are characterized by the largest errors. The latter are generally observed for large $Pe$ (advection-dominated conditions), where gradients become sharper.

\begin{figure*}
	\centering
	\includegraphics[width=1\textwidth]{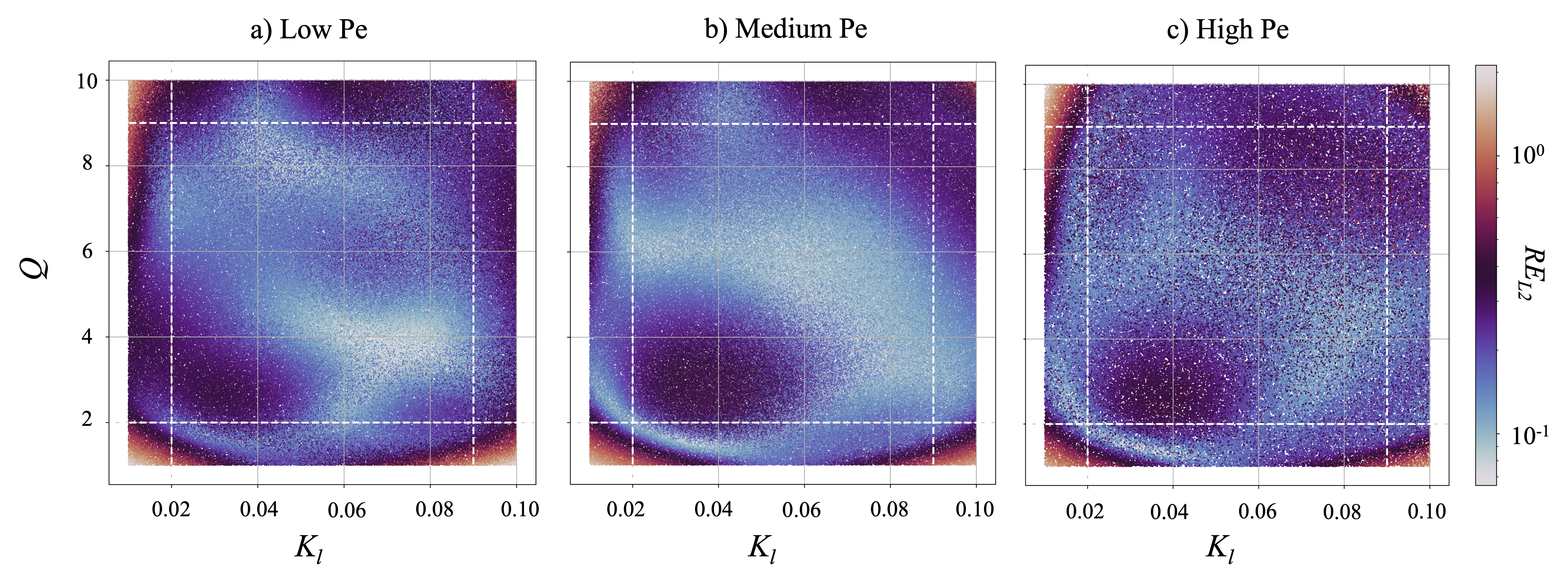}
	\caption{PINN-UU blind spots. Relative percentage errors $RE_{L2}$ (Eq. \eqref{RE}) between the reference FDM solution and the examined PINN-UU architectures in the $K_l$ and $Q$ parameter space. Results are evaluated on the basis of $10^6$ Monte Carlo parameter combination samples and correspond to a) Low, b) Medium, and c) High P\'eclet.}
	\label{Error regions}
\end{figure*}

Notably, in the worst-performing case (Fig. \ref{Error regions}) PINN-UU only slightly underestimates the reference solution in steep regions (advection fronts). When considering the impact of parameter values on model performance (see Table \ref{table4}), we observe that the worst-performing models are overall associated with $K_l$ and $Q$ values close to the boundary of their corresponding intervals. This suggests that low accuracy regions (here termed blind spots) may exist near the boundaries of the parameter space $\boldsymbol{\lambda}_u$. Figure \ref{Error regions} confirms the existence of these blind spots, depicting the distribution of $RE_{L2}$ with regards to parameters $K_l$ and $Q$.

As stated in Section \ref{NN}, training points are selected through Sobol sampling, a quasi-Monte Carlo technique, which is employed to efficiently sample the vast parameter space ($\Omega_{\boldsymbol{\lambda}}$) of our problem formulation.
We recall that the distribution of samples ensuing the application of Sobol sampling can still be influenced by various factors, including the dimensionality of the parameter space and the specific characteristics of the problem. The so-called blind spots can be attributed to both the boundary effects, where system behavior can change significantly near parameter boundaries, and the aforementioned under-sampling issue. The latter arises because the distribution of samples may not adequately capture the complexities of these critical regions. As such, while our model demonstrates a good performance in well-sampled areas, its accuracy diminishes near parameter boundaries, likely due to the limited sample density available for training. Note also that, as opposed to the physical time-space domain, no boundary condition can be assigned along the $\boldsymbol{\lambda}_u$ of the domain $\Omega_{\boldsymbol{\lambda}}$, this factor possibly contributing to the observed accuracy loss. Understanding these limitations is critical for interpreting the reliability of PINN-UU results in scenarios that approach the boundaries of the parameter space. To mitigate this challenge, future investigations may consider augmenting sampling density near boundaries or exploring alternative sampling strategies to enhance model performance across the entire parameter space.
\begin{figure*}
	\centering
	\includegraphics[width=1\textwidth]{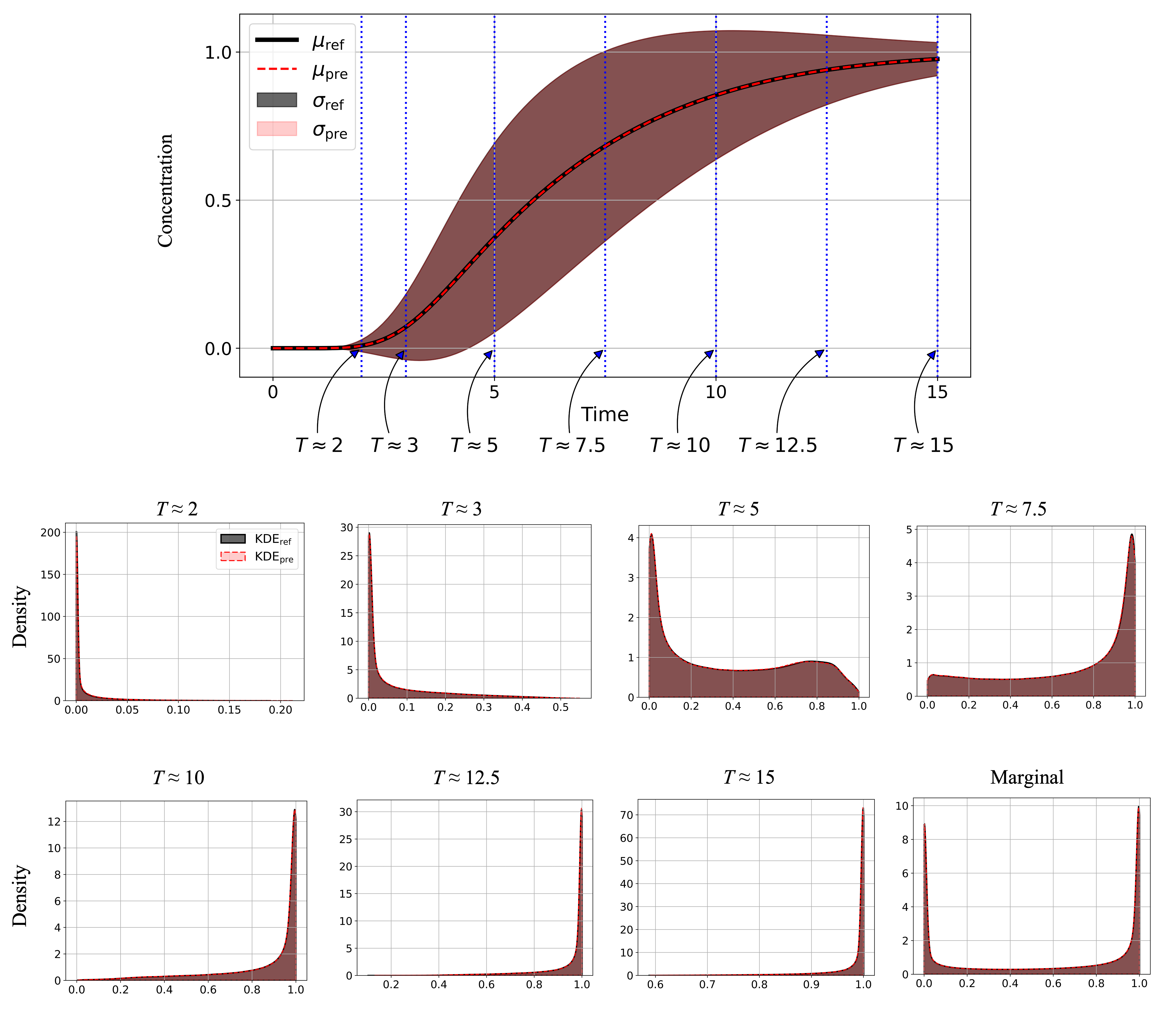}
	\caption{Statistical analysis of solute breakthrough curves (BTCs) samples. The upper panel shows the mean (red dashed and black solid curves) and the standard deviation (red and black combined shadowed areas) of the temporal evolution of effluent concentration ($u_e$) obtained through $10^6$ Monte Carlo samples from the parameter space generated by our PINN-UU compared to reference numerical solver (FDM). The remaining panels illustrate sample PDFs of effluent concentrations conditional on selected time (as indicated by black arrows in the first panel). The marginal (with respect to time) probability distribution of the effluent concentration, representing the probability density of attaining a given outlet concentration value independent of time, is also depicted. Note that the shaded areas with diverse colors essentially overlap in all panels.}
	\label{Probability}
\end{figure*}

\subsubsection{Statistical Analysis of Model Outputs}\label{Statistical wise}
To gain further insights into the predictive capabilities of our PINN-UU framework, in Figure \ref{Probability} we juxtapose the temporal evolution of mean ($\mu$) and standard deviation ($\sigma$) of the effluent concentrations, $u_e$, computed on the bases of $10^6$ Monte Carlo simulation using PINN-UU ($\mu_\text{pre}$ and $\sigma_\text{pre}$, the corresponding mean and standard deviation, respectively) and FDM ($\mu_\text{ref}$ and $\sigma_\text{ref}$, the corresponding mean and standard deviation, respectively). Note that the FDM model requires performing a complete simulation for each sampling point. Otherwise, once our PINN-UU is trained, it produces the results on-demand post-training. Figure \ref{Probability} also includes a Kernel Density Estimations (KDE) of the probability density functions of effluent concentrations at selected times (i.e., the PDF conditional to time). The marginal (with respect to time) probability distribution of the effluent concentration, representing the probability density of attaining a given outlet concentration value independent of time, is also depicted. Note that time level $T = 3$ corresponds to one pore volume, i.e., the median solute arrival time that would be attained without sorption.

At all times, the first two moments of the concentration as wells as concentration PDFs obtained with PINN-UU essentially overlap with those obtained via FDM. This analysis substantiates the PINN-UU accuracy and provides insights into its ability to capture the underlying uncertainty in the parameter space. Furthermore, we observe that the PDFs of effluent concentration are heavily skewed, shifting from right- to a left-skewed distributions as time increases. These results demonstrate the ability of PINN-UU to provide an accurate propagation of uniform input distributions (of parameters) onto otherwise non-uniform and asymmetric PDFs of solute concentrations.

\begin{figure*}
	\centering
	\includegraphics[width=1\textwidth]{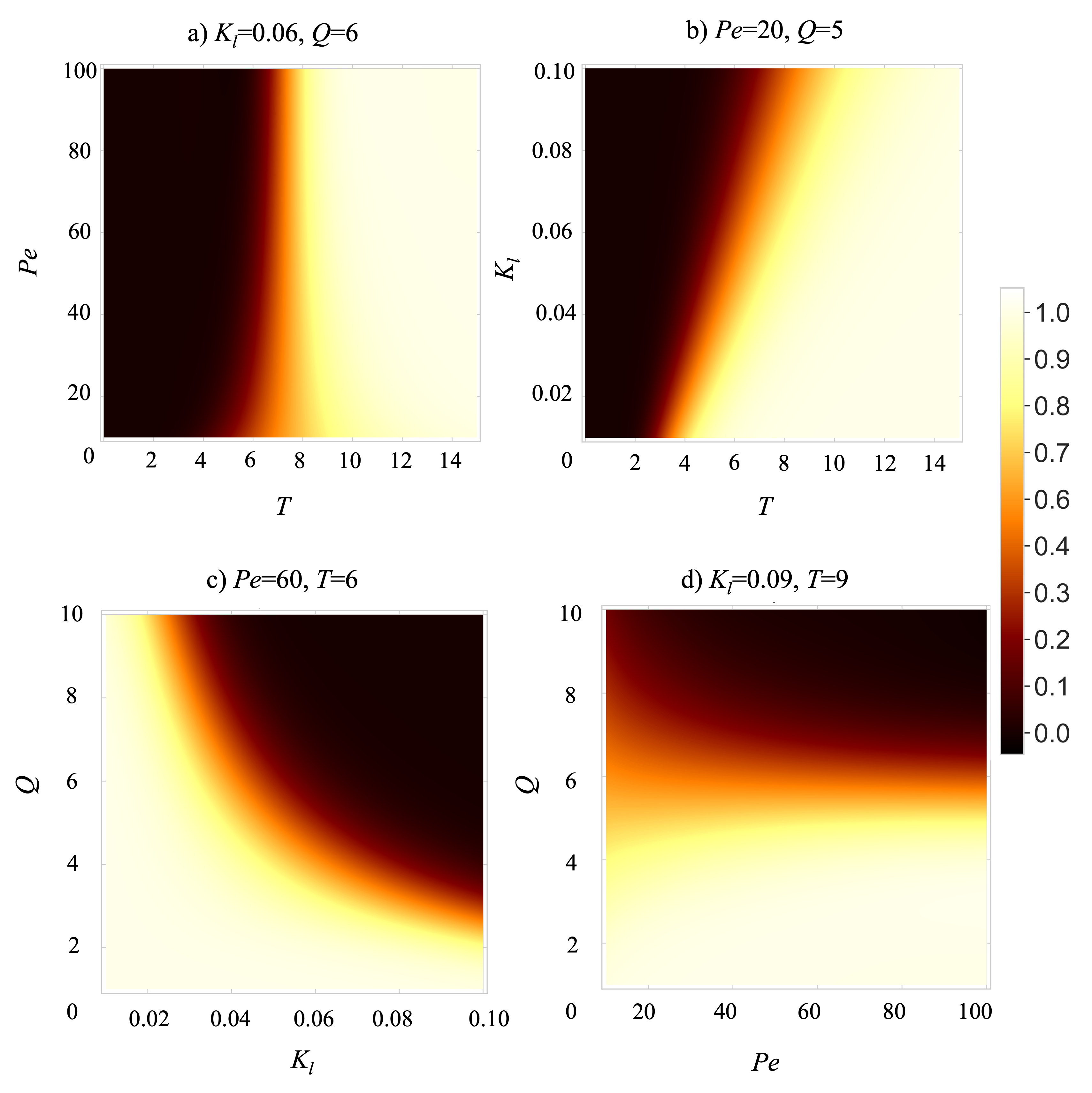}
	\caption{Impact of different combinations of time, $T$, and $[D, K_l, Q]$ on solute concentrations at the system outlet.}
	\label{Solution Variation}
\end{figure*}

\subsection{Sensitivity Analysis (SA)}\label{SA}
Here, we assess the applicability of our PINN-UU model in the context of sensitivity analysis. Typically, Global Sensitivity Analysis (GSA) and Local Sensitivity Analysis (LSA) frameworks require multiple model realizations. PINN-UU infers on-demand the required sensitivity indices through simple and inexpensive post-processing steps. Figure \ref{Solution Variation} provides examples of the types of PINN-UU inferences one can obtain documenting the variability of the effluent solute concentration $u_e$ across the support of two parameters (i.e., $[D, K_l, Q]$), the remaining parameters remaining fixed. In the following sections, we explicitly explore the applicability of our model in the contexts of GSA (Section \ref{GSA}) and LSA (Section \ref{LSA}). Our Sensitivity Analysis (SA) starts by considering the performance of the transport system as rendered through our target quantity, i.e., solute concentration at the system outlet ($u_e$).
\begin{figure*}
	\centering
	\includegraphics[width=1\textwidth]{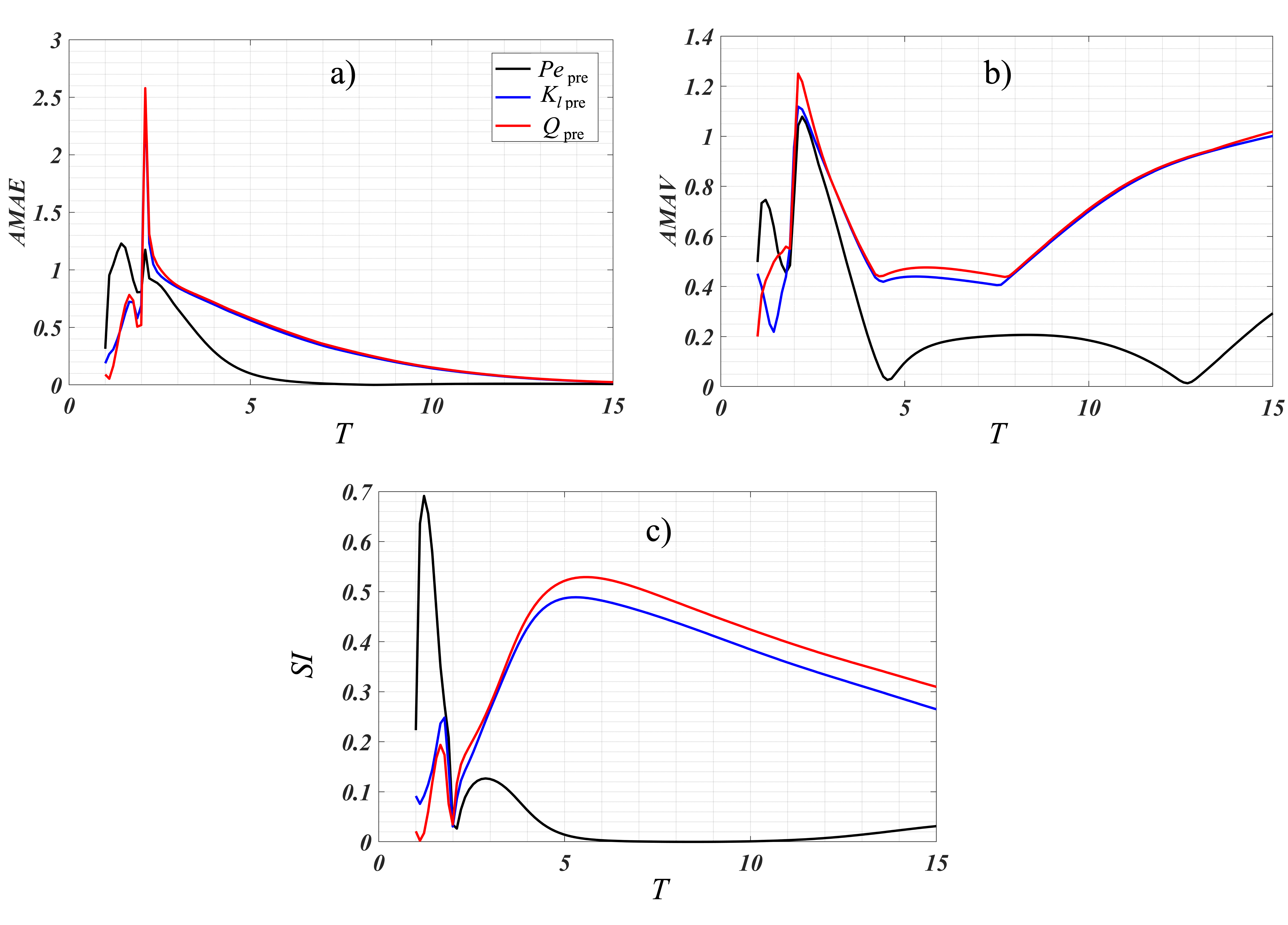}
	\caption{Temporal distributions of the 3 global sensitivity indices considered for solute concentrations at the system outlet ($u_e$), i.e., a) $AMAE$, b) $AMAV$, and c) principal Sobol Indices ($SI$) for three model parameters (P\'eclet number ($Pe$) in black; Adsorption rate ($K_l$) in blue; and Maximum adsorption ($Q$) in red).}
	\label{GSAs}
\end{figure*}
\subsubsection{Global Sensitivity Analysis}\label{GSA}
Our PINN-UU model is characterized by a clear potential for comprehensive parameter space exploration in the context of global sensitivity analyses. Our exemplary analyses consider three typically used global sensitivity metrics, i.e., the variance-based Sobol indices \citep{sobol1993sensitivity} and the moment-based AMA indices (Dell’Oca et al., 2017).
The principal Sobol index represents the relative expected reduction of process variance due to knowledge of (or conditioning on) a model parameter as:

\begin{equation}
SI_{\lambda_i}=\frac{V\left[E\left[u_e\mid \lambda_i\right]\right]}{V[u_e]}=\frac{\left[V[u_e ]-E\left[V\left[u_e  \mid \lambda_i\right]\right]\right]
}{V[u_e ]},
\label{SI}
\end{equation}
where $E[.]$ and $V[.]$ refer to expected value and variance, respectively.

\cite{DellOca2017} propose GSA metrics (termed AMA indices) based on the first four (statistical) moments of the PDF of target model outputs. This technique provides enhanced information content to enrich the commonly used variance-based GSA \citep{sobol1993sensitivity}, because uncertainty and sensitivity are characterized by investigating multiple (statistical) moments of the probability density function of the model outputs, rather than being confined to the variance only. For the purpose of our exemplary study, we consider GSA as rendered through the assessment of the following AMA metrics:
\begin{equation}
AMAE_{\lambda_i}^{u_e } =\left\{\begin{array}{c}
\frac{1}{\left|E[u_e]\right|} E\left[\left| E[u_e]-E\left[u_e  \mid \lambda_i\right]\right|\right], \quad \text{if } E[u_e] \neq 0\\\\
E\left[\left|E\left[u_e  \mid \lambda_i\right]\right|\right], \quad \quad \quad \quad \quad  \text{ if } E[u_e] = 0
\end{array}\right.
\label{AMAE}
\end{equation}

\begin{equation}
AMAV_{\lambda_i}^{ u_e } =\frac{\left.E\left[\mid V\left[u_e \right]-V\left[u_e  \mid \lambda_i\right]\right]\right]}{V\left[u_e \right]},
\label{AMAV}
\end{equation}

Here, $AMAE_{\lambda_i}^{ u_e }$, and $AMAV_{\lambda_i}^{ u_e }$ represent the sensitivity indices associated with the first (mean) and second (variance) moments of $ u_e $, respectively, as linked to variability of $\lambda_i$; the symbol $ u_e \mid \lambda_i$ indicates conditioning of $ u_e $ to a known value of $\lambda_i$ within $\Lambda_{\lambda_i}$. Indices $AMAE_{\lambda_i}^{ u_e }$ and $AMAV_{\lambda_i}^{ u_e }$ respectively quantify the expected change of the mean and variance of $ u_e $ due to our knowledge of $\lambda_i$.

Figure \ref{GSAs} depicts the temporal evolution of the three GSA indices described above with respect to all uncertain model parameters. We note some oscillations in the values of the $AMAE$ and $AMAV$ indices at early times. These could be related to the behavior of the PINN-UU near the outlet boundary region during the initial time steps. Concentration attains values very close to zero in these early stages, thus reflecting the influence of the initial condition. We observe that the parameters related to the sorption model have higher influence on the solution than the dispersion coefficient (here embedded in $Pe$), particularly for $T>3$. The indices related to the two parameters $K_l$ and $Q$ essentially overlap. We also observe that index $AMAV$ reveals mild sensitivity of $ u_e $ to $Pe$ for $T>5$, an effect that is not captured through $S_i$. Otherwise, the value of $Pe$ does not impact the average value of $ u_e $ for $T>5$ (as seen by the virtually zero values for $AMAE$ in Fig. \ref{GSAs}). Having at our disposal this kind of information can be critical when diagnosing a model behavior through sensitivity analysis and to assess parameter identifiability and estimation in an inverse modeling framework. The PINN-UU formulation enables extensive sampling at low computational cost, which is a key requirement to efficiently estimate sensitivity indices for high-order moments.
\begin{figure*}
	\centering
	\includegraphics[width=1\textwidth]{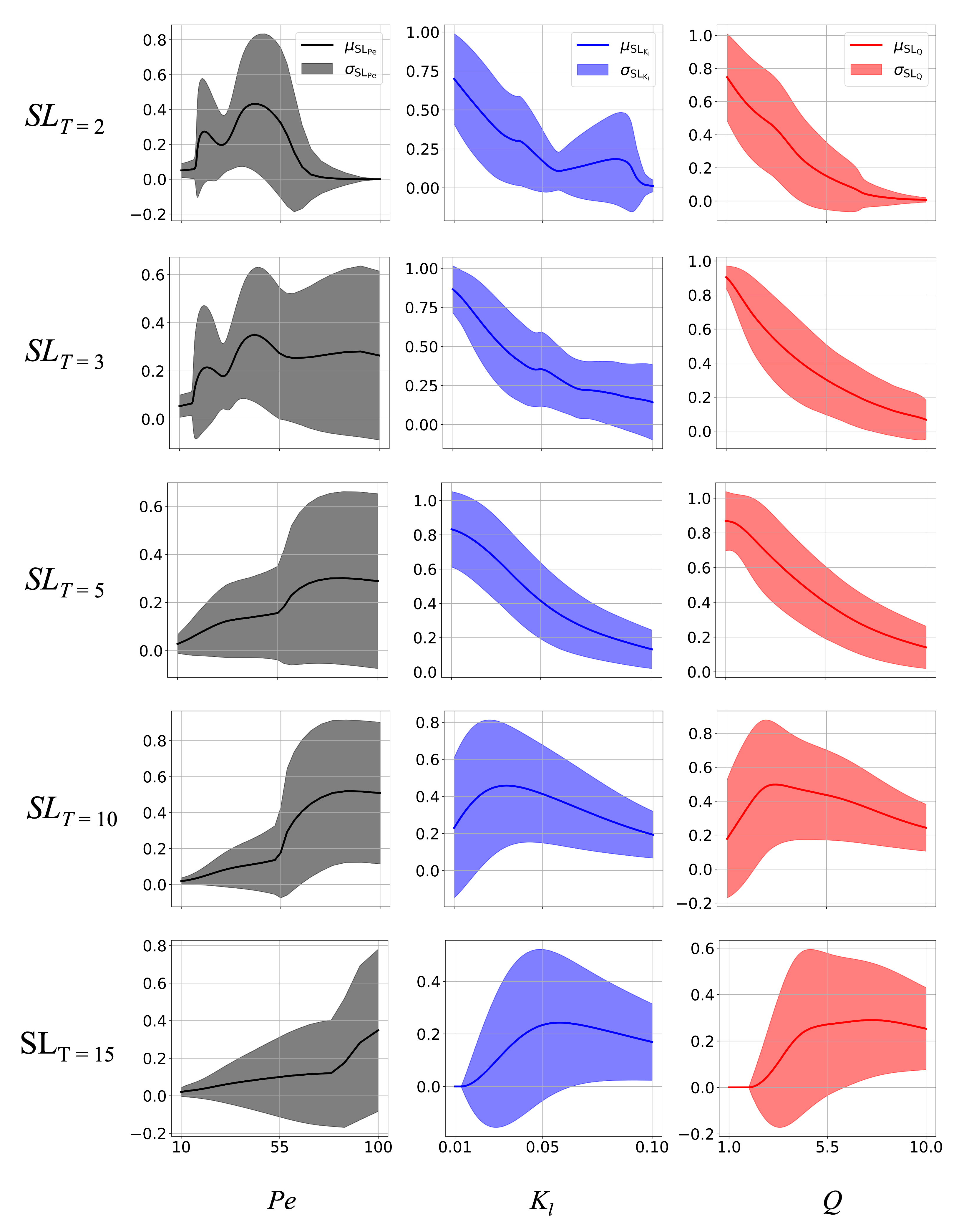}
	\caption{Local Sensitivity Indices. Variability of first two statistical moments of local sensitivity indices (mean $\mu_{SL_{\boldsymbol{\lambda}}}$, and standard deviation $\sigma_{SL_{\boldsymbol{\lambda}}}$) with model parameters, i.e., P\'eclet number ($Pe$; in black), Adsorption rate ($K_l$; in blue), and Maximum adsorption ($Q$; in red) as calculated through Distributed Evaluation of Local Sensitivity Analysis ($DESLA$) at the outlet section of the system for times $T = \{2 , 3, 5, 10, 15\}$.}
	\label{LSAs1}
\end{figure*}
\subsubsection{Local Sensitivity Analysis}\label{LSA}
Local Sensitivity Approaches are typically based on the computation of local (i.e., corresponding to a specific location in the parameter space) derivatives of the model output with respect to parameter values. To reveal how the sensitivity of a model output varies across the parameter space, \cite{rakovec2014distributed} suggest performing a Distributed Evaluation of Local Sensitivity Analysis ($DELSA$). The latter essentially consists of performing multiple evaluations of LSA across the parameter space to evaluate the local sensitivity indices $S_{L_i}$ for each uncertain parameter as follows:
\begin{equation}
    \left\{\begin{array}{l}
S^{j}_{L_i} = \frac{\left(\left.\frac{\partial u_e}{\partial \lambda_i}\right|_{p^j}\right)^2 V[\lambda_i]}{V^{j}_L(u_e)} \quad , \quad 1 \leq i \leq 3\\\\
V_L^{j}(u_e) = \sum_{i=1}^{3} \left(\left.\frac{\partial u_e}{\partial \lambda_i}\right|_{p^j}\right)^2 V[\lambda_i],
\end{array}\right.
\label{eq:delsa}
\end{equation}
where $V[.]$ refers to variance and $p^j$ refers to the grid structure nodes of the uniformly partitioned three-dimensional parameter support $\boldsymbol{\lambda}_u$ (i.e., the uncertain parameters space), partial derivatives ($\frac{\partial u_e}{\partial \lambda_i}$) being evaluated at such nodes. Here, each parameter support $\Lambda_{\lambda_i}$ is partitioned into $100$ intervals. Thus, we employ $1$ million points in the parameter space considering the above mentioned three uncertain parameters (i.e., \( j = 1, 2, \ldots, 1 \times 10^6 \)).

Figure \ref{LSAs1} depicts sensitivity maps (in terms of mean and standard deviation) of local sensitivity indices $S_{L_i}$ as a function of model parameter values, for $5$ time steps. Figure \ref{LSAs1} allows identifying regions of the parameter space characterized by diverse responses to input parameter variability. Notably, effluent concentration, $u_e$, displays similar sensitivity patterns to $K_l$ and $Q$, these two parameters likely having complementary effects on solute breakthrough times (see also Figure \ref{Solution Variation}c). Overall, the importance of $Pe$ is smaller if compared with the sorption parameters, this result being consistent with the GSA results (see Figure \ref{GSAs}). However, we observe that local sensitivities to $Pe$ generally tends to increase for $Pe > 55$. We recall that the mesh-free partial derivatives of solute concentrations at the system outlet ($u_e$) with respect to the parameters are inferred almost instantly as a direct post-processing of our PINN-UU. The terms $\frac{\partial u_e}{\partial \lambda_i}$ appearing in \eqref{eq:delsa} can then be computed without having to replace these derivatives with incremental ratios, as routinely done when employing standard numerical solvers. The PINN-UU neural network allows computation of such gradient through automatic differentiation, the same approach employed for spatial and temporal derivatives.

\section{Conclusions}
\label{conclusions_section}
We develop a physics-informed neural network under uncertainty (PINN-UU) to bridge between physics-informed optimization and uncertainty quantification. We show how our PINN-UU based neural solver can take uncertain PDE parameters as input to its fully physics-driven network. Key conclusions stemming from results are listed in the following.

\begin{enumerate}
  \item Our work documents how to embed the multi-dimensional space associated with uncertain parameters of a PDE taken to describe the dynamics of a physical system within the conceptual and operational framework associated with PINNs. We introduce a step-wise training procedure upon leveraging on transfer learning (warm-starting) techniques (Sections \ref{Transfer learning} and \ref{NN}). The procedure enables us to:\\
  \begin{enumerate}
    \item alleviate the complexities related to the tuning of the parameters of the neural solver in high-dimensional problems, where the choice of the prior distributions for the physical and neural network parameters does not have a significant impact on the approximation performance anymore,
    \item ease computational costs of the training process by an optimum initialization technique.\\
  \end{enumerate}
  \item We rely on unsupervised learning to infer the solution over multiple instances of a parametric nonlinear PDE. Our numerical results are related to a transport scenario featuring advection-dispersion coupled with nonlinear sorption and closely correspond to conditions documented in laboratory scale column experiments. When compared to a benchmark numerical solution, as rendered through a traditional Finite Difference solution of the governing PDE, PINN-UU results have acceptable generalization errors even in advection dominated cases.
  \item We test the capability of PINN-UU in accurately propagating uncertainty from input parameters to model outputs, such as solute breakthrough curves. Once trained, PINN-UU can yield probability distributions of target model outputs through inexpensive sampling of the solution space. PINN-UU can reproduce the probability distributions of temporally varying concentrations at the system outlet with high degree of fidelity when compared with the results of a standard Monte Carlo simulation performed through the benchmark Finite Difference solution. The model is capable of accurately evaluating markedly skewed probability distributions of solute concentrations featuring marked changes in their shape along time.
  \item Outputs of PINN-UU are amenable to be readily employed in the context of global and local sensitivity analysis. After training, the model can provide solution maps as a function of space, time, and parameters at negligible additional computational cost. These can then be readily exploited to evaluate sensitivity metrics, such as moment-based global sensitivity indices (Section \ref{GSA}) or the DELSA indicators (Section \ref{LSA}). Our proposed method can then assist to appropriately design the parametric search space in field/laboratory scale sampling campaigns.
\end{enumerate}

Our approach shows promising potential in addressing high dimensional transient linear/non-linear PDEs. Most of the modifications of vanilla PINNs, such as, for example, bounded residual-based attention (RBA) \citep{anagnostopoulos2023residual}, locally adaptive activation functions \citep{jagtap2020locally}, and gradient enhanced PINNs \citep{yu2022gradient}, to name but a few, can be implemented within the PINN-UU framework. In this context, addressing solute transport within randomly heterogeneous porous media represents a clear future direction for further developments of PINN-UU.

\section*{Declarations}
\label{section_declaration}

\textit{Funding.} The authors acknowledge financial support from the European Union’s Horizon 2020 research and innovation programme under the Marie Skłodowska-Curie grant agreement No 956384\\\\
\textit{Code availability.} Codes used in this paper will be available in the following github repository \url{https://miladpnh.github.io}

\bibliography{References}


\end{document}